# Determining the Dirac CP violation phase and neutrino mass hierarchy

Zoran Bozidar Todorovic

Faculty of Electrical Engineering, Department of Physics, University of Belgrade, Belgrade, Serbia

Email address:
tzoran221@gmail.com

**Abstract:** There is still a problem in neutrino physics related to the configuration of neutrino masses: Are neutrinos arranged by masses following the Standard Model as three generations of fundamental particles, Gen III>Gen II>Gen I, thus forming a structural-normal hierarchy, or deviate from that principle? The biggest obstacle that is still present is the sign of the absolute value of the difference of the square of neutrino masses. It was avoided by applying the theory of neutrino oscillation probability for each structure of the neutrino mass hierarchy. With such theoretical approach the equation of motion was derived for each structure in which Dirac CP violation phase appeared as an unknown quantity. This enables direct calculation of the explicit value for the Dirac CP violation phase. Two examples were analyzed: The first example is devoted to the normal mass ordering and the second one is devoted to the inverted mass ordering. The data used for theoretical calculations presented in this paper are obtained on the basis of the latest reassessed data by processing the results of experimental measurements. On the basis of the performed calculations, normal mass ordering is unconditionally excluded as a potential option regarding the neutrino mass ordering in nature. On the basis of the derived equation of neutrino motion, a possible numerical value of the Dirac CP violation phase and Jarlskog invariant is found.

**Keywords:** Special Relativity, Leptons, Ordinary Neutrino, Neutrino Mass and Mixing, PMNS Matrix, Jarlskog Invariant, Dirac CP Phase

PACS numbers: 03.30.+p,14.60.-z, 14.60.Lm, 14.60.Pq, 11.25.Wx,12.60.Jv

## 1 Introduction

The nature of neutrinos related to neutrino flavor oscillations was experimentally resolved [1,2,3,4], which made it clear that neutrinos could possess mass. The entire theory on the oscillations of neutrino flavor states is based on the application of the unitary PMNS mixing matrix containing parameters which connect flavor eigenstates with mass eigenstates. Those parameters provide mismatches between flavor states and mass eigenstates that are necessary for establishing oscillations between certain neutrino flavor states.

In the standard scenario, the three neutrinos $\nu_1, \nu_2, \nu_3$ are known to have relative masses measured as $\Delta m_{21}^2 = m_2^2 - m_1^2$ and $\Delta m_{31}^2 = |m_3^2 - m_1^2|$. In neutrino physics, the sign of $\Delta m_{31}^2$ is still debatable and unspecified as it has not been measured yet, and that allows two different configurations for the masses: either $m_1 < m_2 < m_3$ (normal mass ordering) or $m_3 < m_1 < m_2$ (inverted mass ordering). The dilemmas in neutrino physics related to the sign of $\Delta m_{31}^2$ still present an obstacle for defining the neutrino mass ordering. However, we will see that the precise definition of the mass splittings between neutrino mass eigenstates, which is done in the latest analysis of experimental data [10], can be of crucial importance for defining the nature of neutrino mass hierarchy. The Standard Model has three generations of fundamental matter particles. Three generations of the quark and charged lepton mass show a hierarchical structure: Gen III > Gen II > Gen I. Owing to that, there is a belief and it is considered that neutrinos may follow such hierarchical structure. Thus, a justified question is raised: Does neutrino mass show the same structure?

And, owing to that and in relation to that, it should be noted that there are more open questions in neutrino physics that are waiting to be answered [1,2,3,4,5,6]:

1. Are squared neutrino masses ordered normally $m_1^2 < m_2^2 < m_3^2$ or are they inverted $m_3^2 < m_1^2 < m_2^2$?
2. What is the numerical value of the Dirac CP violation phase?
3. What is the numerical value of the Jarlskog invariant?
4. What is the lowest neutrino mass? What is the absolute mass of a neutrino?
5. Do neutrinos and antineutrinos behave differently? Is a neutrino its own antiparticle?

In the researches presented in this paper, we will deal with the matters related to the listed points 1, 2, and 3.

## 2 Defining basic relations in Neutrino Physics

Let the wavelengths of oscillations be denoted by $L_{ij}(i,j=1,2,3)$, linking them to the differences of the appropriate phases $\phi_{ij}(i,j=1,2,3)$, and then relations for the processes of disappearances can be written as follows:

**Normal mass ordering,** $m_1 < m_2 < m_3$

$$(\nu_e \to \nu_\mu \to \nu_e) \to \phi_1 - \phi_2 = \phi_{12} = \frac{L_{12}}{\hbar}(p_1 - p_2) = 2\pi$$

$$(\nu_e \to \nu_\tau \to \nu_e) \to \phi_1 - \phi_3 = \phi_{13} = \frac{L_{13}}{\hbar}(p_1 - p_3) = 2\pi,$$

$$(\nu_\mu \to \nu_\tau \to \nu_\mu) \to \phi_2 - \phi_3 = \phi_{23} = \frac{L_{23}}{\hbar}(p_2 - p_3) = 2\pi$$

(1)

The first relation presents the process of oscillation of the electron neutrino through muon neutrino when one full oscillation is performed $L_{12}$.

The second relation presents the process of oscillation of the electron neutrino through tau neutrino when one full oscillation is performed $L_{13}$.

The third relation presents the process of oscillation of the muon neutrino through tau neutrino when one full oscillation is performed $L_{23}$.

The momentum $p_1$ is linked to mass eigenstate $m_1$, the momentum $p_2$ is linked to mass eigenstate $m_2$, the momentum $p_3$ is linked to mass eigenstate $m_3$. From these equations (1), the link between the wavelengths of oscillations is obtained and the corresponding difference of the momentums with the Planck constant:

$$L_{12}(p_1 - p_2) = h$$
(2)
$$L_{13}(p_1 - p_3) = h$$
(3)
$$L_{23}(p_2 - p_3) = h$$
(4)

Where it can be seen that the product of wavelengths of neutrino oscillations and corresponding differences of the momentums equals the Planck constant.

From these equations, a link between wavelengths of oscillations for normal mass ordering (NO) is obtained:

$$\frac{1}{L_{13}} = \frac{1}{L_{12}} + \frac{1}{L_{23}}; L_{12} > L_{23} > L_{13}$$
(5)

In further research, we form the differences of phases of mass eigenstates on the distance $X = L_{12}$ from the source of the neutrino beam, moving through a physical vacuum, and they can be described by following equations:

$$\phi_{12} = \frac{L_{12}}{\hbar}(p_1 - p_2) = \frac{L_{12}}{\hbar}[E/c(1-\delta_1) - E/c(1-\delta_2)] = \frac{L_{12}}{\hbar}[E/c(\delta_2 - \delta_1)]$$
$$= \frac{L_{12}}{\hbar}\frac{E}{c}\left(\frac{m_2^2 c^4}{2E^2} - \frac{m_1^2 c^4}{2E^2}\right) = \frac{L_{12} c^3}{2\hbar E}(m_2^2 - m_1^2) = \frac{L_{12} c^3}{2\hbar E}\Delta m_{21}^2 = 2\pi$$

$$m_3 > m_2 > m_1; \delta_1 = \frac{m_1^2 c^4}{2E^2} \ll 1, \delta_2 = \frac{m_2^2 c^4}{2E^2} \ll 1, \delta_3 = \frac{m_3^2 c^4}{2E^2} \ll 1.$$
(6)

$$\phi_{23} = \frac{L_{12}}{\hbar}(p_2 - p_3) = \frac{L_{12}}{\hbar}[E/c(\delta_3 - \delta_2)] = \frac{L_{12} c^3}{2\hbar E}\Delta m_{23}^2 = 2\pi \frac{\Delta m_{32}^2}{\Delta m_{21}^2}.$$
(7)

$$\phi_{13} = \frac{L_{12}}{\hbar}(p_1 - p_3) = \frac{L_{12}}{\hbar}[E/c(\delta_3 - \delta_1)] = \frac{L_{12} c^3}{2\hbar E}\Delta m_{31}^2 = 2\pi \frac{\Delta m_{31}^2}{\Delta m_{21}^2}$$
(8)

where $c$ is the speed of light, and $\hbar = h/2\pi$ and one more equation can be written:

$$\Delta m_{21}^2 + \Delta m_{32}^2 = \Delta m_{31}^2$$
(9)

**Inverted mass ordering,** $m_3 < m_1 < m_2$

$$(\nu_e \to \nu_\mu \to \nu_e) \to \phi_1 - \phi_2 = \phi_{12} = \frac{L_{12}}{\hbar}(p_1 - p_2) = 2\pi$$

$$(\nu_e \to \nu_\tau \to \nu_e) \to \phi_3 - \phi_1 = \phi_{31} = \frac{L_{31}}{\hbar}(p_3 - p_1) = 2\pi,$$

$$(\nu_\mu \to \nu_\tau \to \nu_\mu) \to \phi_3 - \phi_2 = \phi_{32} = \frac{L_{32}}{\hbar}(p_3 - p_2) = 2\pi.$$
(10)

The first relation describes the process of oscillations of the electron neutrino through muon neutrino when one full oscillation is performed $L_{12}$.

The second relation presents the process of oscillations of the electron neutrino through tau neutrino when one full oscillation is performed $L_{31}$.

The third relation presents the process of oscillations of the muon neutrino through tau neutrino when one full oscillation is performed $L_{32}$.

The momentum $p_1$ is linked to the mass eigenstate $m_1$, the momentum $p_2$ is linked to the mass eigenstate $m_2$, the momentum $p_3$ is linked to the mass eigenstate $m_3$. The equations signify that the product of wavelengths of neutrino oscillations and corresponding differences of the momentums equals the Planck constant.

From the relations (10), the following equations directly follow:

$$L_{12}(p_1 - p_2) = h$$
(11)
$$L_{31}(p_3 - p_1) = h$$
(12)
$$L_{32}(p_3 - p_2) = h$$
(13)

From which it can be seen that the product of wavelengths of neutrino oscillations and corresponding differences of the momentums equals the Planck constant $h$.

From these equations, the link between wavelengths of oscillations for inverted mass ordering (IMO) is obtained:

$$\frac{1}{L_{32}} = \frac{1}{L_{12}} + \frac{1}{L_{31}}; L_{32} < L_{31} < L_{12}$$
(14)

Since wavelengths of oscillations are directly proportional to the neutrino energy, these relations apply to any neutrino energy, and they change in proportion to the energy, which should be taken into account when this relation is applied.

Phase differences of mass eigenstates on the distance $X = L_{12}$ from the source of the neutrino beam, moving through a vacuum, can be described by following equations:

$$\phi_{12} = \frac{L_{12}}{\hbar}(p_1 - p_2) = \frac{L_{12}}{\hbar}\left[E/c(\delta_2 - \delta_1)\right] = \frac{L_{12}c^3}{2\hbar E}\Delta m_{21}^2 = 2\pi$$
$$m_3 < m_1 < m_2; \delta_1 = \frac{m_1^2 c^4}{2E^2} \ll 1, \delta_2 = \frac{m_2^2 c^4}{2E^2} \ll 1, \delta_3 = \frac{m_3^2 c^4}{2E^2} \ll 1.$$
(15)

$$\phi_{32} = \frac{L_{12}}{\hbar}(p_3 - p_2) = \frac{L_{12}}{\hbar}\left[E/c(\delta_2 - \delta_3)\right] = \frac{L_{12}c^3}{2\hbar E}\Delta m_{23}^2 = 2\pi\frac{\Delta m_{23}^2}{\Delta m_{21}^2}$$
(16)

$$\phi_{31} = \frac{L_{12}}{\hbar}(p_3 - p_1) = \frac{L_{12}}{\hbar}\left[E/c(\delta_1 - \delta_3)\right] = \frac{L_{12}c^3}{2\hbar E}\Delta m_{13}^2 = 2\pi\frac{\Delta m_{13}^2}{\Delta m_{21}^2}$$
(17)

where $c$ is the speed of light, and $\hbar = h/2\pi$ and

$$\Delta m_{21}^2 + \Delta m_{13}^2 = \Delta m_{23}^2$$
(18)

## 3 Application of the $U_{PMNS}^{PDG}$ mixing matrix

In the processes known as neutrino flavor oscillations, the Dirac CP violation phase $\delta_{CP}$ is unequivocally singled out as the cause of those oscillations in the propagation of the neutrino beam through the physical vacuum. For that reason, there arises the question of writing the equation of motion in which $\delta_{CP}$ would appear as an unknown quantity. On the basis of that equation, it would be possible to determine that unknown quantity. So far, there appears to be only one way to derive equations of motion for a neutrino beam, and it is related to the use of the equations of the neutrino oscillations probabilities. The procedure for deriving those equations is given here.

### 3.1 The case of normal hierarchy of neutrino mass (NO)

In this case, the matrix $U_{PMNS}^{PDG}$ is used [5,8,9,11,12]

$$U_{PMNS}^{PDG} = \begin{pmatrix} U_{e1} & U_{e2} & U_{e3} \\ U_{\mu1} & U_{\mu2} & U_{\mu3} \\ U_{\tau1} & U_{\tau2} & U_{\tau3} \end{pmatrix}$$

$$= \begin{pmatrix} c_{12}c_{13} & s_{12}c_{13} & s_{13}e^{-i\delta_{CP}} \\ -s_{12}c_{23} - c_{12}s_{23}s_{13}e^{i\delta_{CP}} & c_{12}c_{23} - s_{12}s_{23}s_{13}e^{i\delta_{CP}} & c_{13}s_{23} \\ s_{12}s_{23} - c_{12}c_{23}s_{13}e^{i\delta_{CP}} & -c_{12}s_{23} - s_{12}c_{23}s_{13}e^{i\delta_{CP}} & c_{13}c_{23} \end{pmatrix}$$

$$= \begin{pmatrix} U_{e1} & U_{e2} & Je^{-i\delta_{CP}} \\ -A - Be^{i\delta_{CP}} & C - De^{i\delta_{CP}} & U_{\mu3} \\ E - Fe^{i\delta_{CP}} & -G - He^{i\delta_{CP}} & U_{\tau3} \end{pmatrix}$$
(19)

where the mixing angles from the [10] are taken into consideration:

$$\theta_{12} = 34.3 \pm 1^0, \theta_{23} = 49.26 \pm 0.79^0, \theta_{13} = 8.53^{+0.13^0}_{-0.53},$$
$$c_{ij} = \cos\theta_{ij}, s_{ij} = \sin\theta_{ij}; i, j = 1,2,3.$$

In order to obtain an explicit numerical value of $\delta_{CP}$, the following unconditional rule will be applied: The sum of the probabilities of three neutrino oscillations during the transition $\nu_e \to \nu_\mu, \nu_e \to \nu_\tau, \nu_e \to \nu_e$, at a distance from the source equal to the entire wavelength of oscillations in the value of $X = L_{12}$, during the process of the disappearance in transition $\nu_e \to \nu_\mu \to \nu_e$, in the propagation of the neutrino beam through vacuum (as it can be seen, the matter effect is excluded in these considerations), is equal to one.

Ref. [10] provides the following data as well:

$$\Delta m_{21}^2 = 7.50^{+0.22}_{-0.20} \times 10^{-5} eV^2, \Delta m_{31}^2 = 2.55^{+0.02}_{-0.03} \times 10^{-3} eV^2, \Delta m_{32}^2 = 2.475 \times 10^{-3}$$

(20)

Taking into consideration the central values from the data (20), we get:

$$\frac{\Delta m_{31}^2}{\Delta m_{21}^2} = \frac{0.00255}{0.0000750} = 34.0, \frac{\Delta m_{32}^2}{\Delta m_{21}^2} = \frac{0.002475}{0.0000750} = 33.0,$$

$$V = \sin\left(2\pi \frac{\Delta m_{31}^2}{\Delta m_{21}^2}\right) = 0, V = \sin\left(2\pi \frac{\Delta m_{32}^2}{\Delta m_{21}^2}\right) = 0,$$

$$W = \sin^2\left(\pi \frac{\Delta m_{31}^2}{\Delta m_{21}^2}\right) = 0, W = \sin^2\left(\pi \frac{\Delta m_{32}^2}{\Delta m_{21}^2}\right) = 0.$$

(21)

### Neutrino motion equation

In our considerations, we will use the general formula for neutrino oscillations given in Ref [8,9]:

$$P(\nu_\alpha \to \nu_\beta) = \delta_{\alpha\beta} - 4\sum_{i,<j} R\left(U_{\alpha i}U^*_{\beta i}U^*_{\alpha j}U_{\beta j}\right)\sin^2\frac{\Delta m_{ji}^2 Lc^3}{4E\hbar}$$

$$+ 2\sum_{i,<j} \text{Im}\left(U_{\alpha i}U^*_{\beta i}U^*_{\alpha j}U_{\beta j}\right)\sin\frac{\Delta m_{ji}^2 Lc^3}{2E\hbar}; i,j = 1,2,3.$$

(22)

**The transition is analysed:** $\nu_e \to \nu_\mu, \nu_e \to \nu_\tau, \nu_e \to \nu_e$

On the basis of formulae (22), the total probability of neutrino oscillations is shown through the equation

$$P(\nu_e \to \nu_\mu) + P(\nu_e \to \nu_\tau) + P(\nu_e \to \nu_e)$$

$$= 1 - 4R\left\{U_{e1}U^*_{\mu 1}U^*_{e2}U_{\mu 2}\sin^2\pi\frac{\Delta m_{21}^2}{\Delta m_{21}^2}\right\} + 2\text{Im}\left\{U_{e1}U^*_{\mu 1}U^*_{e2}U_{\mu 2}\sin 2\pi\frac{\Delta m_{21}^2}{\Delta m_{21}^2}\right\}$$

$$- 4R\left\{U_{e1}U^*_{\mu 1}U^*_{e3}U_{\mu 3}\sin^2\pi\frac{\Delta m_{31}^2}{\Delta m_{21}^2}\right\} + 2\text{Im}\left\{U_{e1}U^*_{\mu 1}U^*_{e3}U_{\mu 3}\sin 2\pi\frac{\Delta m_{31}^2}{\Delta m_{21}^2}\right\}$$

$$- 4R\left\{U_{e2}U^*_{\mu 2}U^*_{e3}U_{\mu 3}\sin^2\pi\frac{\Delta m_{32}^2}{\Delta m_{21}^2}\right\} + 2\text{Im}\left\{U_{e2}U^*_{\mu 2}U^*_{e3}U_{\mu 3}\sin 2\pi\frac{\Delta m_{32}^2}{\Delta m_{21}^2}\right\}$$

$$- 4R\left\{U_{e1}U^*_{\tau 1}U^*_{e2}U_{\tau 2}\sin^2\pi\frac{\Delta m_{21}^2}{\Delta m_{21}^2}\right\} + 2\text{Im}\left\{U_{e1}U^*_{\tau 1}U^*_{e2}U_{\tau 2}\sin 2\pi\frac{\Delta m_{21}^2}{\Delta m_{21}^2}\right\}$$

$$- 4R\left\{U_{e1}U^*_{\tau 1}U^*_{e3}U_{\tau 3}\sin^2\pi\frac{\Delta m_{31}^2}{\Delta m_{21}^2}\right\} + 2\text{Im}\left\{U_{e1}U^*_{\tau 1}U^*_{e3}U_{\tau 3}\sin 2\pi\frac{\Delta m_{31}^2}{\Delta m_{21}^2}\right\}$$

$$- 4R\left\{U_{e2}U^*_{\tau 2}U^*_{e3}U_{\tau 3}\sin^2\pi\frac{\Delta m_{32}^2}{\Delta m_{21}^2}\right\} + 2\text{Im}\left\{U_{e2}U^*_{\tau 2}U^*_{e3}U_{\tau 3}\sin 2\pi\frac{\Delta m_{32}^2}{\Delta m_{21}^2}\right\}$$

$$- 4|U_{e1}|^2|U_{e2}|^2 \sin^2 \pi \frac{\Delta m_{21}^2}{\Delta m_{21}^2} - 4|U_{e1}|^2|U_{e3}|^2 \sin^2 \pi \frac{\Delta m_{31}^2}{\Delta m_{21}^2}$$

$$- 4|U_{e2}|^2|U_{e3}|^2 \sin^2 \pi \frac{\Delta m_{32}^2}{\Delta m_{21}^2} = 1$$

(23)

And, from the equation (23), the equation of neutrino motion is formed with a condition that the travelled distance of the neutrino beam, moving through a vacuum from the source, equals the neutrino wavelength $X = L_{12}$. So, it can be written as

$$-4R\left\{U_{e1}U_{\mu1}^*U_{e2}^*U_{\mu2}\sin^2\pi\frac{\Delta m_{21}^2}{\Delta m_{21}^2}\right\}+2\operatorname{Im}\left\{U_{e1}U_{\mu1}^*U_{e2}^*U_{\mu2}\sin 2\pi\frac{\Delta m_{21}^2}{\Delta m_{21}^2}\right\}$$

$$-4R\left\{U_{e1}U_{\mu1}^*U_{e3}^*U_{\mu3}\sin^2\pi\frac{\Delta m_{31}^2}{\Delta m_{21}^2}\right\}+2\operatorname{Im}\left\{U_{e1}U_{\mu1}^*U_{e3}^*U_{\mu3}\sin 2\pi\frac{\Delta m_{31}^2}{\Delta m_{21}^2}\right\}$$

$$-4R\left\{U_{e2}U_{\mu2}^*U_{e3}^*U_{\mu3}\sin^2\pi\frac{\Delta m_{32}^2}{\Delta m_{21}^2}\right\}+2\operatorname{Im}\left\{U_{e2}U_{\mu2}^*U_{e3}^*U_{\mu3}\sin 2\pi\frac{\Delta m_{32}^2}{\Delta m_{21}^2}\right\}$$

$$-4R\left\{U_{e1}U_{\tau1}^*U_{e2}^*U_{\tau2}\sin^2\pi\frac{\Delta m_{21}^2}{\Delta m_{21}^2}\right\}+2\operatorname{Im}\left\{U_{e1}U_{\tau1}^*U_{e2}^*U_{\tau2}\sin 2\pi\frac{\Delta m_{21}^2}{\Delta m_{21}^2}\right\}$$

$$-4R\left\{U_{e1}U_{\tau1}^*U_{e3}^*U_{\tau3}\sin^2\pi\frac{\Delta m_{31}^2}{\Delta m_{21}^2}\right\}+2\operatorname{Im}\left\{U_{e1}U_{\tau1}^*U_{e3}^*U_{\tau3}\sin 2\pi\frac{\Delta m_{31}^2}{\Delta m_{21}^2}\right\}$$

$$-4R\left\{U_{e2}U_{\tau2}^*U_{e3}^*U_{\tau3}\sin^2\pi\frac{\Delta m_{32}^2}{\Delta m_{21}^2}\right\}+2\operatorname{Im}\left\{U_{e2}U_{\tau2}^*U_{e3}^*U_{\tau3}\sin 2\pi\frac{\Delta m_{32}^2}{\Delta m_{21}^2}\right\}$$

$$-4|U_{e1}|^2|U_{e2}|^2\sin^2\pi\frac{\Delta m_{21}^2}{\Delta m_{21}^2}-4|U_{e1}|^2|U_{e3}|^2\sin^2\pi\frac{\Delta m_{31}^2}{\Delta m_{21}^2}$$

$$-4|U_{e2}|^2|U_{e3}|^2\sin^2\pi\frac{\Delta m_{32}^2}{\Delta m_{21}^2}=0$$

(24)

Inserting data (21), it can be seen that all members of the equation (24), with no exception, become equal to zero, and that it is always the case regardless of the values of the elements of PMNS mixing matrix (19) and their connections with the Dirac CP violation phase.

The complex structure of equations (24) is reduced to an extremely simple form:

$$(4W\cos\delta_{CP}-2V\sin\delta_{CP})*0=(4\cos\delta_{CP}*0-2\sin\delta_{CP}*0)*0=0$$

(25)

Every member of the equation (25) equals zero. That means that CP phases can have any arbitrarily taken value from the interval $(0,2\pi)$. This result can be considered an irrefutable proof that, in nature, neutrinos, in the hierarchy of the mass eigenstates, do not belong to the normal mass ordering, $m_1 < m_2 < m_3$. From this, the only conclusion that follows is that, in nature, the neutrino mass eigenstates belong to the inverted mass ordering, $m_3 < m_1 < m_2$ to which we shall pay further attention.

### 3.2 The case of inverted hierarchy of neutrino masses (IO)

And, in this case, the matrix $U_{PMNS}^{PDG}$ is used [5,8,9,11,12]

$$U_{PMNS}^{PDG}=\begin{pmatrix}U_{e1}&U_{e2}&U_{e3}\\U_{\mu1}&U_{\mu2}&U_{\mu3}\\U_{\tau1}&U_{\tau2}&U_{\tau3}\end{pmatrix}$$

$$=\begin{pmatrix}c_{12}c_{13}&s_{12}c_{13}&s_{13}e^{-i\delta_{CP}}\\-s_{12}c_{23}-c_{12}s_{23}s_{13}e^{i\delta_{CP}}&c_{12}c_{23}-s_{12}s_{23}s_{13}e^{i\delta_{CP}}&c_{13}s_{23}\\s_{12}s_{23}-c_{12}c_{23}s_{13}e^{i\delta_{CP}}&-c_{12}s_{23}-s_{12}c_{23}s_{13}e^{i\delta_{CP}}&c_{13}c_{23}\end{pmatrix}$$

$$=\begin{pmatrix}U_{e1}&U_{e2}&Je^{-i\delta_{CP}}\\-A-Be^{i\delta_{CP}}&C-De^{i\delta_{CP}}&U_{\mu3}\\E-Fe^{i\delta_{CP}}&-G-He^{i\delta_{CP}}&U_{\tau3}\end{pmatrix}$$

(26)

For determining values of PMNS matrix elements, as well as for defining elements of the motion equation, the data given in the Ref. [10] are used:

$$\theta_{12} = 34.3 \pm 1^0, \theta_{23} = 49.46 \pm 0.79^0, \theta_{13} = 8.58^{+0.13^0}_{-0.53}, c_{ij} = \cos\theta_{ij}, s_{ij} = \sin\theta_{ij};$$
$$i, j = 1,2,3..$$

(27)
$$\Delta m^2_{21} = 7.50^{+0.22}_{-0.20} \times 10^{-5} eV^2, \Delta m^2_{13} = 2.45^{+0.02}_{-0.03} \times 10^{-3} eV^2, \Delta m^2_{23} = 2.525 \times 10^{-3} eV^2$$

To obtain an explicit numerical value of $\delta_{CP}$, the following unconditional rule will be applied: The sum of the probabilities of three neutrino oscillations during the transition $\nu_e \to \nu_\mu, \nu_e \to \nu_\tau, \nu_e \to \nu_e$, at a distance from the source equal to the wavelength of oscillations in the value of $X = L_{12}$. For this transition, total oscillation probabilities equal one:

$$P(\nu_e \to \nu_\mu) + P(\nu_e \to \nu_\tau) + P(\nu_e \to \nu_e)$$
$$= 1 - 4R\left\{U_{e1}U^*_{\mu1}U^*_{e2}U_{\mu2} \sin^2 \pi \frac{\Delta m^2_{21}}{\Delta m^2_{21}}\right\} + 2\text{Im}\left\{U_{e1}U^*_{\mu1}U^*_{e2}U_{\mu2} \sin 2\pi \frac{\Delta m^2_{21}}{\Delta m^2_{21}}\right\}$$
$$- 4R\left\{U_{e1}U^*_{\mu1}U^*_{e3}U_{\mu3} \sin^2 \pi \frac{\Delta m^2_{31}}{\Delta m^2_{21}}\right\} + 2\text{Im}\left\{U_{e1}U^*_{\mu1}U^*_{e3}U_{\mu3} \sin 2\pi \frac{\Delta m^2_{31}}{\Delta m^2_{21}}\right\}$$
$$- 4R\left\{U_{e2}U^*_{\mu2}U^*_{e3}U_{\mu3} \sin^2 \pi \frac{\Delta m^2_{32}}{\Delta m^2_{21}}\right\} + 2\text{Im}\left\{U_{e2}U^*_{\mu2}U^*_{e3}U_{\mu3} \sin 2\pi \frac{\Delta m^2_{32}}{\Delta m^2_{21}}\right\}$$
$$- 4R\left\{U_{e1}U^*_{\tau1}U^*_{e2}U_{\tau2} \sin^2 \pi \frac{\Delta m^2_{21}}{\Delta m^2_{21}}\right\} + 2\text{Im}\left\{U_{e1}U^*_{\tau1}U^*_{e2}U_{\tau2} \sin 2\pi \frac{\Delta m^2_{21}}{\Delta m^2_{21}}\right\}$$
$$- 4R\left\{U_{e1}U^*_{\tau1}U^*_{e3}U_{\tau3} \sin^2 \pi \frac{\Delta m^2_{31}}{\Delta m^2_{21}}\right\} + 2\text{Im}\left\{U_{e1}U^*_{\tau1}U^*_{e3}U_{\tau3} \sin 2\pi \frac{\Delta m^2_{31}}{\Delta m^2_{21}}\right\}$$
$$- 4R\left\{U_{e2}U^*_{\tau2}U^*_{e3}U_{\tau3} \sin^2 \pi \frac{\Delta m^2_{32}}{\Delta m^2_{21}}\right\} + 2\text{Im}\left\{U_{e2}U^*_{\tau2}U^*_{e3}U_{\tau3} \sin 2\pi \frac{\Delta m^2_{32}}{\Delta m^2_{21}}\right\}$$
$$- 4|U_{e1}|^2|U_{e2}|^2 \sin^2 \pi \frac{\Delta m^2_{21}}{\Delta m^2_{21}} - 4|U_{e1}|^2|U_{e3}|^2 \sin^2 \pi \frac{\Delta m^2_{31}}{\Delta m^2_{21}} - 4|U_{e2}|^2|U_{e3}|^2 \sin^2 \pi \frac{\Delta m^2_{32}}{\Delta m^2_{21}} = 1$$

(28)

From the equation (28), the following form of the motion equation ensues:

$$-4R\left\{U_{e1}U_{\mu1}^{*}U_{e2}^{*}U_{\mu2}\sin^{2}\pi\frac{\Delta m_{21}^{2}}{\Delta m_{21}^{2}}\right\}+2\text{Im}\left\{U_{e1}U_{\mu1}^{*}U_{e2}^{*}U_{\mu2}\sin 2\pi\frac{\Delta m_{21}^{2}}{\Delta m_{21}^{2}}\right\}$$

$$-4R\left\{U_{e1}U_{\mu1}^{*}U_{e3}^{*}U_{\mu3}\sin^{2}\pi\frac{\Delta m_{31}^{2}}{\Delta m_{21}^{2}}\right\}+2\text{Im}\left\{U_{e1}U_{\mu1}^{*}U_{e3}^{*}U_{\mu3}\sin 2\pi\frac{\Delta m_{31}^{2}}{\Delta m_{21}^{2}}\right\}$$

$$-4R\left\{U_{e2}U_{\mu2}^{*}U_{e3}^{*}U_{\mu3}\sin^{2}\pi\frac{\Delta m_{32}^{2}}{\Delta m_{21}^{2}}\right\}+2\text{Im}\left\{U_{e2}U_{\mu2}^{*}U_{e3}^{*}U_{\mu3}\sin 2\pi\frac{\Delta m_{32}^{2}}{\Delta m_{21}^{2}}\right\}$$

$$-4R\left\{U_{e1}U_{\tau1}^{*}U_{e2}^{*}U_{\tau2}\sin^{2}\pi\frac{\Delta m_{21}^{2}}{\Delta m_{21}^{2}}\right\}+2\text{Im}\left\{U_{e1}U_{\tau1}^{*}U_{e2}^{*}U_{\tau2}\sin 2\pi\frac{\Delta m_{21}^{2}}{\Delta m_{21}^{2}}\right\}$$

$$-4R\left\{U_{e1}U_{\tau1}^{*}U_{e3}^{*}U_{\tau3}\sin^{2}\pi\frac{\Delta m_{31}^{2}}{\Delta m_{21}^{2}}\right\}+2\text{Im}\left\{U_{e1}U_{\tau1}^{*}U_{e3}^{*}U_{\tau3}\sin 2\pi\frac{\Delta m_{31}^{2}}{\Delta m_{21}^{2}}\right\}$$

$$-4R\left\{U_{e2}U_{\tau2}^{*}U_{e3}^{*}U_{\tau3}\sin^{2}\pi\frac{\Delta m_{32}^{2}}{\Delta m_{21}^{2}}\right\}+2\text{Im}\left\{U_{e2}U_{\tau2}^{*}U_{e3}^{*}U_{\tau3}\sin 2\pi\frac{\Delta m_{32}^{2}}{\Delta m_{21}^{2}}\right\}$$

$$-4|U_{e1}|^{2}|U_{e2}|^{2}\sin^{2}\pi\frac{\Delta m_{21}^{2}}{\Delta m_{21}^{2}}-4|U_{e1}|^{2}|U_{e3}|^{2}\sin^{2}\pi\frac{\Delta m_{31}^{2}}{\Delta m_{21}^{2}}-4|U_{e2}|^{2}|U_{e3}|^{2}\sin^{2}\pi\frac{\Delta m_{32}^{2}}{\Delta m_{21}^{2}}=0$$

(29)

Where

$$\sin\left(\pi\frac{\Delta m_{13}^{2}}{\Delta m_{21}^{2}}\right)=\sin\left(\pi\frac{24500}{750}\right)=+\frac{\sqrt{3}}{2}$$

(30)

$$\sin\left(\pi\frac{\Delta m_{23}^{2}}{\Delta m_{21}^{2}}\right)=\sin\left(\pi\frac{25250}{750}\right)=-\frac{\sqrt{3}}{2}$$

(31)

$$W=\sin^{2}\left(\pi\frac{\Delta m_{13}^{2}}{\Delta m_{21}^{2}}\right)=\frac{3}{4}$$

(32)

$$W=\sin^{2}\left(\pi\frac{\Delta m_{23}^{2}}{\Delta m_{21}^{2}}\right)=\frac{3}{4}$$

(33)

$$V=\sin\left(2\pi\frac{\Delta m_{13}^{2}}{\Delta m_{21}^{2}}\right)=\sin\left(2\pi\frac{24500}{750}\right)=-\frac{\sqrt{3}}{2}$$

(34)

$$V=\sin\left(2\pi\frac{\Delta m_{23}^{2}}{\Delta m_{21}^{2}}\right)=\sin\left(2\pi\frac{25250}{750}\right)=-\frac{\sqrt{3}}{2}$$

(35)

Since

$$V=\sin\left(2\pi\frac{\Delta m_{13}^{2}}{\Delta m_{21}^{2}}\right)=\sin\left(2\pi\frac{\Delta m_{23}^{2}}{\Delta m_{21}^{2}}\right), W=\sin^{2}\left(\pi\frac{\Delta m_{13}^{2}}{\Delta m_{21}^{2}}\right)=\sin^{2}\left(\pi\frac{\Delta m_{23}^{2}}{\Delta m_{21}^{2}}\right)=\frac{3}{4}$$

(36)

The equation (29) is reduced to the form:

$$3J[U_{e1}(AU_{\mu3}-EU_{\tau3})-U_{e2}(CU_{\mu3}-GU_{\tau3})]\cos\delta_{CP}$$
$$+2VJ[U_{e1}(U_{\tau3}E-U_{\mu3}A)+U_{e2}(U_{\mu3}C-U_{\tau3}G)]\sin\delta_{CP}$$
$$-3U_{e1}^{2}J^{2}-3U_{e2}^{2}J^{2}+3J(BU_{\mu3}+FU_{\tau3})+3J(DU_{\mu3}+HU_{\tau3})$$
$$=(3J\cos\delta_{CP}-2VJ\sin\delta_{CP})(U_{e1}AU_{\mu3}-U_{e1}U_{\tau3}E+U_{e2}CU_{\mu3}-U_{e2}GU_{\tau3})$$
$$-3U_{e1}^{2}J^{2}-3U_{e2}^{2}J^{2}+3JU_{e1}(BU_{\mu3}+FU_{\tau3})+3JU_{e2}(DU_{\mu3}+HU_{\tau3})=0$$

(37)

Or in a simplified form

$$(3J\cos\delta_{CP} - 2VJ\sin\delta_{CP})\varsigma - \xi = 0$$
(38)

In this equation, the following expressions equal zero:

$$\varsigma = (U_{e1}AU_{\mu3} - U_{e1}EU_{\tau3} - U_{e2}CU_{\mu3} + U_{e2}GU_{\tau3}) = 0$$
(39)

Because

$$U_{\mu3}A - U_{\tau3}E = S_{23}C_{13} \times S_{12}C_{23} - C_{23}C_{13} \times S_{12}S_{23} = 0$$
(40)

$$U_{\tau3}G - U_{\mu3}C = C_{23}C_{13} \times C_{12}S_{23} - S_{23}C_{13} \times C_{12}S_{23} = 0$$
(41)

$$\xi = -3U_{e1}^2 J^2 - 3U_{e2}^2 J^2 + 3U_{e1}J(BU_{\mu3} + FU_{\tau3}) + 3U_{e2}J(DU_{\mu3} + HU_{\tau3}) = 0$$

(42)

Because

$$U_{\mu3}B + U_{\tau3}F - U_{e1}J = S_{23}C_{13} \times C_{12}S_{23}S_{13} + C_{23}C_{13} \times C_{12}C_{23}S_{13} - C_{12}C_{13}S_{13} = 0$$

(43)

$$U_{\mu3}D + U_{\tau3}H - U_{e2}J = S_{23}C_{13} \times S_{12}S_{23}S_{13} + C_{23}C_{13} \times S_{12}C_{23}S_{13} - S_{12}C_{13}S_{13} = 0$$

(44)

Additionally, when the appropriate data from the experimental measurements are included [10], the equation (38) is reduced to the form:

$$(3\cos\delta_{CP} - 2V\sin\delta_{CP}) * 0 = 0$$
(45)

The first point that can be stated is that this equation is always satisfied for any value of $\delta_{CP} \in [0, 2\pi)$, so such solutions make no physical sense. It is apparent that among those solutions in the range $\delta_{CP} \in [0, 2\pi)$ there is the right unique solution for the value $\delta_{CP}$. From such set of countless values, the real and unique value for $\delta_{CP}$ is drawn from the set $\delta_{CP} \in [0, 2\pi)$ by solving the equation

$$3\cos\delta_{CP} - 2V\sin\delta_{CP} = 0$$
(46)

The solution of this equation presents the particular solution of the equation (45), and it is in the following form:

$$tg\delta_{CP} = \frac{3}{2V} = \frac{3}{2\sin\left(2\pi\frac{\Delta m_{13}^2}{\Delta m_{21}^2}\right)} = \frac{3}{2\sin\left(2\pi\frac{\Delta m_{23}^2}{\Delta m_{21}^2}\right)} = \frac{3}{2\left(-\frac{\sqrt{3}}{2}\right)} = -\sqrt{3}$$

(47)

It is especially significant to emphasize the existence of the common factor (39) in the equation (38) and it equals zero. That is why the equation could be written in the form (45). However, as we have seen, the structure of the equation for normal mass ordering (NMO) had a different form:

$$(4J\cos\delta_{CP} *W - 2VJ\sin\delta_{CP}*V)*0$$
$$= (4J\cos\delta_{CP}*0 - 2VJ\sin\delta_{CP}*0)*0 = 0; W \neq V; W = 0, V = 0.$$
(48)

And, it, as such, has no sense, because the members in parentheses also equal zero. Using the procedure for calculating total probability, both for normal mass ordering (NMO) and inverted mass ordering (IMO), the same results are obtained. Namely, the total probability for normal mass ordering (23)

$$P(v_e \to v_\mu) + P(v_e \to v_\tau) + P(v_e \to v_e) = 1$$
(49)

In addition, by equating the equation (45) with zero, the total probability in the case of inverted mass hierarchy (28) becomes

$$P(v_e \to v_\mu) + P(v_e \to v_\tau) + P(v_e \to v_e) = 1$$
(50)

The first point that we have established is that neutrinos belong to the inverted mass hierarchy. Secondly, for that hierarchy, we can calculate the numerical value, on the basis of (47), for the Dirac CP violation phase

$$\delta_{CP} = arctg(-\sqrt{3}) = -60^0 = +300^0$$
(51)

The obtained formula for the Dirac CP violation phase (47) exclusively depends on the mass splittings between the neutrino mass eigenstates in a manner established by the experiments [1,2,3,4] so that neutrino oscillations depend only on the mass splittings between the neutrino mass eigenstates.

## 4 Some properties of the parameters of the Dirac CP violation phase

If we go back to the relations provided in the formulae (47), it can be seen that the formula for the Dirac CP violation phase can be expressed through the equivalent formula

$$tg\delta_{CP} = \frac{3}{2V} = \frac{3}{2\sin\phi_{31}} = \frac{3}{2\sin\phi_{32}}$$
(52)

It can also be written that

$$tg\delta_{CP} = \sin\phi_{31} + \sin\phi_{32}$$
(53)

As well as

$$tg\delta_{CP} = \frac{\sin\delta_{CP}}{\cos\delta_{CP}} = \frac{3}{2V} = \frac{3}{2\sin\delta_{CP}}$$
(54)

And, from here

$$\cos\delta_{CP} = \frac{2}{3}\sin^2\delta_{CP} = \frac{1}{2}$$
(55)

Returning to equations (16) and (17), it can be written:

$$\phi_{32} - \phi_{31} = 2\pi \frac{\Delta m^2_{23}}{\Delta m^2_{21}} - 2\pi \frac{\Delta m^2_{13}}{\Delta m^2_{21}} = 2\pi$$

(56)

$$\sin \phi_{32} - \sin \phi_{31} = 0,$$

$$\sin \delta_{CP} = \frac{1}{2}(\sin \phi_{32} + \sin \phi_{31}) = \sin \phi_{32} = \sin \phi_{31}$$

(57)

$$\sin\left(\pi \frac{\Delta m^2_{23}}{\Delta m^2_{21}}\right) = \sin\left(\pi \frac{25250}{750}\right) = -\frac{\sqrt{3}}{2}, \cos\left(\pi \frac{\Delta m^2_{23}}{\Delta m^2_{21}}\right) = \cos\left(\pi \frac{25250}{750}\right) = +\frac{1}{2}, \quad \sin\left(2\pi \frac{\Delta m^2_{13}}{\Delta m^2_{21}}\right) = \sin\left(2\pi \frac{\Delta m^2_{23}}{\Delta m^2_{21}}\right), \sin^2\left(2\pi \frac{\Delta m^2_{13}}{\Delta m^2_{21}}\right) = \sin^2\left(2\pi \frac{\Delta m^2_{23}}{\Delta m^2_{21}}\right) = \frac{3}{4}$$

$$\sin\left(2\pi \frac{\Delta m^2_{23}}{\Delta m^2_{21}}\right) = \sin\left(2\pi \frac{25250}{750}\right) = 2\sin\left(\pi \frac{25250}{750}\right)\cos\left(\pi \frac{25250}{750}\right) = -\frac{\sqrt{3}}{2}, \quad \cos\left(2\pi \frac{\Delta m^2_{13}}{\Delta m^2_{21}}\right) = \cos\left(2\pi \frac{\Delta m^2_{23}}{\Delta m^2_{21}}\right), \cos^2\left(2\pi \frac{\Delta m^2_{13}}{\Delta m^2_{21}}\right) = \cos^2\left(2\pi \frac{\Delta m^2_{23}}{\Delta m^2_{21}}\right) = \frac{1}{4}$$

$$\cos\left(2\pi \frac{\Delta m^2_{23}}{\Delta m^2_{21}}\right) = \cos\left(2\pi \frac{25250}{750}\right) = \cos^2\left(\pi \frac{25250}{750}\right) - \sin^2\left(\pi \frac{25250}{750}\right) = -\frac{1}{2} \quad (60)$$

(58)

$$\sin\left(\pi \frac{\Delta m^2_{13}}{\Delta m^2_{21}}\right) = \sin\left(\pi \frac{24500}{750}\right) = +\frac{\sqrt{3}}{2}, \cos\left(\pi \frac{\Delta m^2_{13}}{\Delta m^2_{21}}\right) = \cos\left(\pi \frac{24500}{750}\right) = -\frac{1}{2},$$

$$\sin\left(2\pi \frac{\Delta m^2_{13}}{\Delta m^2_{21}}\right) = \sin\left(2\pi \frac{24500}{750}\right) = 2\sin\left(\pi \frac{24500}{750}\right)\cos\left(\pi \frac{24500}{750}\right) = -\frac{\sqrt{3}}{2},$$

$$\cos\left(2\pi \frac{\Delta m^2_{13}}{\Delta m^2_{21}}\right) = \cos\left(2\pi \frac{24500}{750}\right) = \cos^2\left(\pi \frac{24500}{750}\right) - \sin^2\left(\pi \frac{24500}{750}\right) = -\frac{1}{2}.$$

(59)

From the relations (58) and (59), it follows:

Moreover, we can calculate the numerical value for Jarlskog invariant [22,23], using for the Dirac CPV phase $\delta_{CP}$ numerical value (51), which is

$$J_{CP(IO)} = J_{CP(IO)}^{\max} \sin \delta_{CP} = s_{12} c_{12} s_{23} c_{23} s_{13} c_{13}^2 \sin \delta_{CP} \approx -0.029$$
(61)

And this value is in the vicinity of the point minimum in the Ref. NuFIT 5.0 (2020), on the CP violation graph: Jarlskog invariant, as it is depicted in Fig.1.

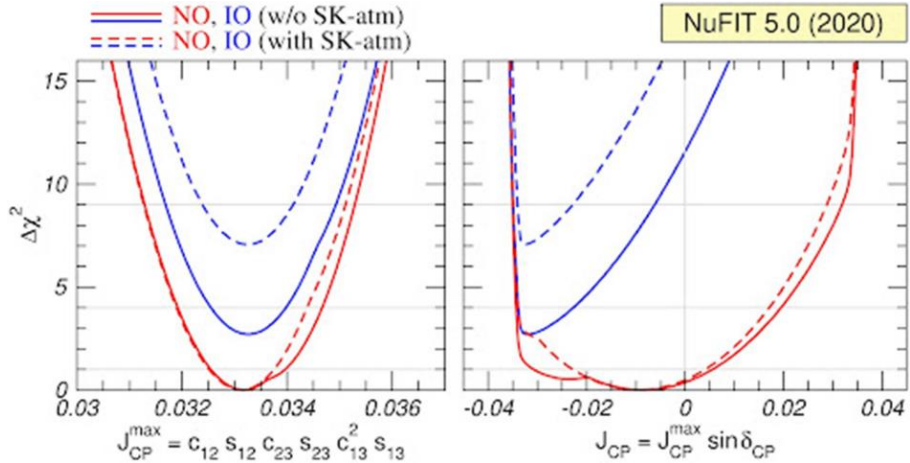

**Figure 1. CP violation: Jarlskog invariant**

## 5 The Results Discussion

The introduced theoretical research is based on the latest data presented in the Ref. [10]. The selection of data processed in that manner is given in the Table 3, depicting neutrino oscillation parameters summary determined from the global analysis. Neutrino oscillations depend only on the mass splittings between the neutrino mass eigenstates. From the Table 3, parameters for both types of ordering are selected:

**Normal ordering**

$$\theta_{12} = 34.3 \pm 1^0, \theta_{23} = 49.26 \pm 0.79^0, \theta_{13} = 8.53^{+0.13^0}_{-0.53},$$
$$c_{ij} = \cos\theta_{ij}, s_{ij} = \sin\theta_{ij}; i,j = 1,2,3.$$
$$\Delta m_{21}^2 = 7.50^{+0.22}_{-0.20} \times 10^{-5} eV^2, \Delta m_{31}^2 = 2.55^{+0.02}_{-0.03} \times 10^{-3} eV^2.$$
(62)

**Inverted ordering**

$$\theta_{12} = 34.3 \pm 1^0, \theta_{23} = 49.46 \pm 0.79^0, \theta_{13} = 8.58^{+0.13^0}_{-0.53},$$
$$c_{ij} = \cos\theta_{ij}, s_{ij} = \sin\theta_{ij}; i,j = 1,2,3.$$
(63)
$$\Delta m_{21}^2 = 7.50^{+0.22}_{-0.20} \times 10^{-5} eV^2, \Delta m_{13}^2 = 2.45^{+0.02}_{-0.03} \times 10^{-3} eV^2$$

Central values from the best fit $\pm 1\sigma$ are taken for calculations:

**Normal ordering**

$$\theta_{12} = 34.3^0, \theta_{23} = 49.26^0, \theta_{13} = 8.53^0, c_{ij} = \cos\theta_{ij}, s_{ij} = \sin\theta_{ij}; i,j = 1,2,3.$$

$$\Delta m_{21}^2 = 7.50 \times 10^{-5} eV^2, \Delta m_{31}^2 = 2.55 \times 10^{-3} eV^2, \Delta m_{32}^2 = 2.475 \times 10^{-3} eV^2$$
(64)

**Inverted ordering**

$$\theta_{12} = 34.3^0, \theta_{23} = 49.46^0, \theta_{13} = 8.58^0, c_{ij} = \cos\theta_{ij}, s_{ij} = \sin\theta_{ij}; i,j = 1,2,3.$$

(65)
$$\Delta m_{21}^2 = 7.50 \times 10^{-5} eV^2, \Delta m_{13}^2 = 2.45 \times 10^{-3} eV^2, \Delta m_{23}^2 = 2.525 \times 10^{-3} eV^2$$

Applying the adopted values for the parameters (64) and (65), we obtained the results (21) and (25) on the basis of which normal mass ordering is disfavoured with no restrictions.
On the basis of the motion equation (29), we obtained the final equation for the Dirac CP violation phase (45), the global solution of which is that there are countless solutions for $\delta_{CP}$, which has no physical sense and must be rejected. However, the equation (46) offers a particular solution for $\delta_{CP}$ and it is essentially singled out from countless solutions offered by the equation (45), but being singled out, it represents a possible solution that makes physical sense for the Dirac CP violation phase (47).
The main characteristic of the solution for CP phase presented by the equation (47) is that it exclusively depends on the mass splittings between the neutrino mass eigenstates.
For the sake of comparison, both obtained numerical values, for CP phase (47) and $\delta_{CP}/\pi \approx 1.667$, are in the vicinity of the local minimum of the inverted ordering, as depicted by the graphs Figure 5 and Figure 8, given in the Ref.[10].

## 6 Conclusions

The purpose of this paper was to obtain an explicit value of the Dirac CP violation phase. Two examples were analyzed: The first example represents normal mass ordering, which is unconditionally excluded due to the structure of equation (25), and which as such does not offer any solution for the CP phase. Furthermore, when it comes to the structure of the neutrino mass hierarchy, it can be determined on the basis of the formula (14). Our research is based on the data obtained on the basis of processing the latest experimental measurements exhibited in the Ref. [10], in which the selection of data processed in that manner is given in the Table 3, depicting neutrino oscillation parameters summary determined from the global analysis.
On the basis of the complex structure of the motion equation (29), we obtained the final simple equation for the Dirac CP violation phase (45) whose particular solution for $\delta_{CP}$ has physical sense and it is given in the formula (47).
It is especially emphasized that the main feature of this solution is that it depends exclusively on the mass splittings between the neutrino mass eigenstates, which is also the main conclusion regarding the existence of the phenomenon of neutrino oscillations based on the mass splittings between neutrino mass eigenstates as shown by the experiments [1,2,3,4,5]. This solution differs from other solutions given in Refs. [16,17,18,19]
The graphs shown in Ref. [10] clearly depict that the obtained numerical values, for CP phase (47) and $\delta_{CP}/\pi \approx 1.667$, are in the vicinity of the local minimum of the inverted ordering, as shown in the graphs Figure 5 and Figure 8, as it is depicted in Fig.2.

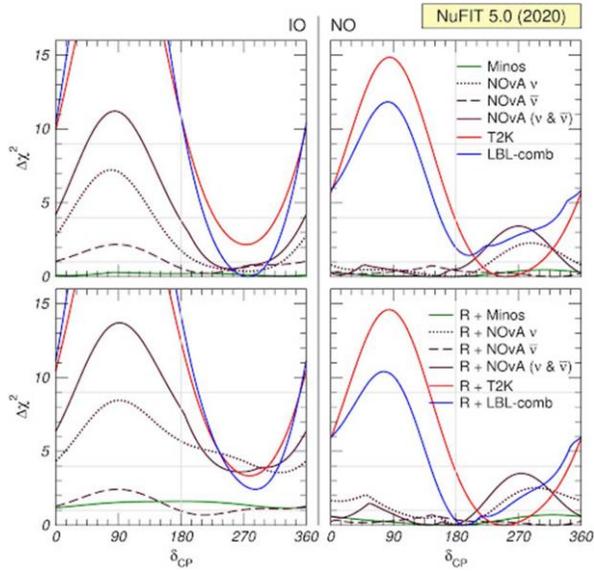

**Figure 2. Determination of $\delta_{CP}$**

Checking the existence of the inverted mass hierarchy could also be easily established through the relation (14). The theoretical results presented in this paper are exclusively related to the data in Ref. [10]. However, if the published data [10] change, that could influence the outcome of the final derived results. Moreover, it is particularly emphasized that there are opinions that neutrino mass hierarchy could have a normal structure [11], but there is no clear and convincing argumentation for that. Additionally, there is an opinion that Ref. [14] does not provide a strong evidence of normal hierarchy over inverted hierarchy. That opinion could be checked through comparison of wavelengths of oscillations given in the formula (5).

Finding a numerical value of the Dirac CP violation phase (51) would enable the calculation of the three neutrino mass eigenstates, as well as the possible effective value of the neutrino Majorana mass, which could appear in the process of the neutrinoless double beta decay [20,21].